\documentclass[preprint,showpacs,preprintnumbers,amsmath,amssymb]{revtex4}
\usepackage{graphicx}
\usepackage{dcolumn}
\usepackage{bm}

\usepackage{color}

\usepackage{color}

\usepackage{color}

\begin{document}

\title{Magnetic properties of small Pt-capped Fe, Co and Ni clusters: A density functional theory study}

\author{Sanjubala Sahoo}
\email{sanjubala.sahoo@uni-due.de}
\affiliation{Physics Department, University of Duisburg--Essen, 
Lotharstr.~1, D-47048 Duisburg, Germany}

\author{Alfred Hucht}
\affiliation{Physics Department, University of Duisburg--Essen, 
Lotharstr.~1, D-47048 Duisburg, Germany}
\author{Markus E. Gruner}
\affiliation{Physics Department, University of Duisburg--Essen, 
Lotharstr.~1, D-47048 Duisburg, Germany}
\author{Peter Entel}
\affiliation{Physics Department, University of Duisburg--Essen, 
Lotharstr.~1, D-47048 Duisburg, Germany}
\author{Andrei Postnikov}
\affiliation{LPMD, Paul Verlaine University -- Metz,
1 Bd Arago, F-57078 Metz, France}
\author{Jaime Ferrer}
\affiliation{Departamento de F\'{\i}sica, Universidad de Oviedo,
33007 Oviedo, Spain}
\author{Lucas Fern\'andez-Seivane}
\affiliation{Departamento de F\'{\i}sica, Universidad de Oviedo,
33007 Oviedo, Spain}
\author{Manuel Richter}
\affiliation{IFW Dresden e.V., Helmholtzstr. 20, P.O. Box 270116,
D-01171 Dresden, Germany}
\author{Daniel Fritsch}
\affiliation{IFW Dresden e.V., Helmholtzstr. 20, P.O. Box 270116,
D-01171 Dresden, Germany}
\author{Shreekantha Sil}
\affiliation{Department of Physics, Visva Bharati University, Santiniketan, 731235 West Bengal, India}
\begin{abstract}
Theoretical studies on M$_{13}$ (M = Fe, Co, Ni) and M$_{13}$Pt$_n$ 
(for $n$ = 3, 4, 5, 20) clusters including the spin-orbit coupling are 
done using density functional theory. The magnetic anisotropy 
energy (MAE) along with the spin and orbital moments are 
calculated for M$_{13}$ icosahedral clusters. The 
angle-dependent energy differences are modelled using an extended classical 
Heisenberg model with local anisotropies.
From our studies, the MAE for Jahn-Teller distorted Fe$_{13}$, Mackay 
distorted Fe$_{13}$ and nearly undistorted Co$_{13}$ clusters are found to be 322, 60 and 
5 $\mu$eV/atom, respectively, and are large 
relative to the corresponding bulk values, (which are 1.4 and 
1.3 $\mu$eV/atom for bcc Fe and fcc Co, respectively.) However, for 
Ni$_{13}$ (which practically does not show relaxation tendencies), 
the calculated value of MAE is found to be 
0.64 $\mu$eV/atom, which is approximately four times smaller compared 
to the bulk fcc Ni (2.7 $\mu$eV/atom). 
In addition, MAE of the capped cluster 
(Fe$_{13}$Pt$_4$) is enhanced 
compared to the uncapped Jahn-Teller distorted Fe$_{13}$ cluster. 
\end{abstract}

\pacs{Valid PACS appear here}
\maketitle

\section{Introduction}
Density-functional theory (DFT) is an adequate tool to obtain useful 
information on the physical properties of small clusters. However, with 
respect to magnetic properties,
the variations of magnetic moments with cluster size and morphology
are often too obscure for establishing a clear trend (for example, see 
the behavior of the exchange coupling of free Fe clusters 
in~\cite{polesya-2006,sipr-2007}). 
Still, certain general statements can be made.  It may be safely concluded 
that the local magnetic moments in outer shells of clusters are enhanced,
as compared to the interior of clusters or to the corresponding bulk
crystal \cite{xie-2002, dieguez-2001, postnikov-2003, duan-2001}. This is a 
typical effect of reduced atomic coordination at the 
surface \cite{apsel-1996}, well known from numerous 
experiments or calculations of magnetic slabs.
Moreover, an analysis of cluster morphologies, by both
experiment and theory \cite{winter-bovin-1991, dunlap-1990, tiago-2006, entel_pm-2008, rollmann-2007, gruner-2008, xiao-2009, futschek-2006}, reveals the importance of icosahedral-like 
structures, prohibited for bulk or 
surface phases. On the other hand, the diversity of cluster 
structures in combination with the surface enhancement of magnetic 
moments make clusters interesting model objects for tuning 
magnetic properties at the nanometer scale. 

When addressing magnetic properties of clusters from the computational
point of view at a realistic level, it is important 
to take into account two specific issues: 
($i$) a possible non-collinear (NC) setting of magnetic moments
(i.e., a smooth variation of magnetic density vector from point
to point in space), and ($ii)$ spin-orbit interaction (SOI), 
along with the existence of orbital moments. Both issues have a long record of 
incorporation into first-principles DFT calculations, and are 
internally related: They mix the spin-up and spin-down states 
and must be, in principle, treated alongside on equal footing
(see, e.g., Refs.~\cite{sandra-1998-NC, sandra-1996-NC-SOI} for a 
review). Discussing specifically DFT calculations for clusters, one 
notes certain technical difficulties in combining NC spin density 
with the SOI, which acted so far as a limiting factor on the number of 
calculations performed, and the size of clusters 
treated: Lack of symmetry, big effect of structure relaxation,
slow convergence, large size of simulation cell around a free cluster. 
Moreover there is a conceptual problem of choosing a ``correct'' 
non-collinear solution among many apparently close metastable 
configurations. Different groups report very different results for the 
same systems (e.g., Refs.~\cite{oda-1998, hobbs-2000, lounis-2007}), so that 
a preference of one or another result is not obvious. Therefore, 
not so much the size of cluster is a problem in itself,  
as the organization of a calculation in a way allowing to extract meaningful 
results, that is, clearly defining structural and magnetic models 
and carefully analyzing their consequences. 

A better understanding of the origins and details of large orbital 
magnetic moment as well as large MAE in 
clusters~\cite{strandberg-2007,gambardella-2003} 
is demanding in order to manipulate materials for improving the 
technology like in magnetic data storage devices. Binary 
$3d$--$5d$ clusters can be a challenging material in this respect. 
For example, there are observations that $4d$ and $5d$ elements
like Rh, Pt and Au, which are nonmagnetic in bulk phase, attain significant 
moments when alloyed with $3d$ transition metals like Fe, 
Co or Ni ~\cite{luis-2006, bansmann-2005, dennler-2004, navia-2004}. 

In the present work, we analyze the effect of capping icosahedral 
clusters M$_{13}$ of M = Fe, Co, and Ni with Pt atoms.
Theoretically, there have been 
several studies on magnetic anisotropy for supported 
clusters ~\cite {masahito-2008, medina-2003, nonas-2001, andriotis-2004, rohart-2006, tsujikawa-2008, pick-2004}, whereas studies on small free clusters 
are still limited~\cite{pastor-1995, ferrer-2007, hong-1991, fritsch-2008, blonski-2009}.
The motivation for the present study is that icosahedral clusters 
for M$_{13}$ are known to be very 
stable,~\cite{guez-2001, rollmann-2007, duan-2001} and the 
alloying of $3d$ transition metals with Pt results in large magnetic 
anisotropy as well as orbital moment. Hence, it is expected that
the orbital and the spin moments of M$_{13}$ clusters be strongly affected 
by capping with Pt.

Most of our DFT calculations have been done with the planewave code 
{\sc Vasp} (Vienna Ab-Initio Simulation Package)~\cite{kresse-cms, kresse-1996, kresse-1999}.
For test purposes and for validating the results of calculations
on non-capped icosahedral clusters, we performed calculations also
with a local-orbitals code, {\sc Siesta} (Spanish Initiative for 
Electronic Simulations with Thousands of Atoms)~\cite{soler-2002}. As 
the two methods are very
different in what regards the technical implementation of the DFT
calculation scheme, and both have been earlier used in calculations 
of magnetic clusters, their direct comparison might be of its own interest.
In addition, few test calculations for the binary clusters were done 
using the all-electron local-orbital code {\sc Fplo} 
(Full-Potential Local-Orbital scheme)~\cite{klaus-1999}. 
In this respect, it is interesting to note that all-electron 
calculations confirm the results obtained with {\sc Vasp} 
and {\sc Siesta}.

The paper is organized as follows. Section~\ref{sec:methods}
outlines the calculational methods and setup. Section III
deals with the results for monometallic icosahedral clusters,
notably comparison between {\sc Vasp} and {\sc Siesta} results. 
Section IV discusses the results for capped clusters,
obtained with the {\sc Vasp} code. We have made a comparison 
between {\sc Vasp} and {\sc Fplo} for one of the capped clusters.  
Conclusions are drawn in Section V.\\

\section{Computational methods}
\label{sec:methods}
Most of the calculations have been performed with the 
{\sc Vasp} code \cite{kresse-cms, kresse-1996, kresse-1999} based on DFT and 
within the generalized gradient approximation (GGA). The parameters by 
Perdew are used for the exchange and 
correlation functional~\cite{perdew-1991}.
{\sc Vasp} uses the projector augmented wave (PAW) 
method \cite{kresse-1999, blochl1994} 
and a planewave basis set. Periodic boundary conditions 
were imposed onto large enough 
cubic cells (with the edge of 15~{\AA} for M$_{13}$ clusters 
and 20~{\AA} for M$_{13}$Pt$_n$ clusters) in order 
to sufficiently reduce the 
interaction between replicated cluster images. Only the $\Gamma$ point 
was used for the Brillouin-zone sampling for the cluster 
calculations. A k-mesh of (11$\times$11$\times$11) is used for the 
bulk calculations to compute the equilibrium lattice constants 
in case of bcc Fe, fcc Ni and Co.
The values for local magnetic moments correspond to the 
integration of continuous magnetization density over atom-centred spheres 
with radii of 1.302~{\AA} (Fe, Co), 1.286~{\AA} (Ni) and 1.455~{\AA} (Pt). 
It is a well known fact that the magnetic anisotropy energy for 
cubic bulk transition metals, defined as the maximum energy difference, 
per atom, between different settings of the spin moment
relative to the atoms framework, is of the order of 10$^{-6}$ eV. 
Moreover, a special care is demanded for the study of the MAE in 
clusters because any slight error can accumulate and produce 
misleading results while dealing 
with energy differences of such a small scale. In order to calculate the 
magnetic anisotropy of M$_{13}$ icosahedral clusters, we have used 
the energy convergence criterion for the self-consistency as 10$^{-10}$ 
eV with a Gaussian half-width parameter of 0.01 eV for 
the discrete energy levels. A very high plane wave cutoff value of 
1000 eV as well as a dense Fourier 
grid spacing of 0.04~\AA~in each of $x$, $y$ and $z$  directions is taken.

While relaxing the clusters using conjugate gradient algorithm, the same 
energy convergence criterion is used but a lower plane wave cut-off of 270 eV. 
The structural relaxation of clusters are done in 
the scalar relativistic mode. For the calculation of orbital 
moment and MAE, the SOI is treated as a perturbation in the 
scalar relativistic Hamiltonian.
Benchmark calculations for some M$_{13}$ clusters have also been
done with the {\sc Siesta} code~\cite{soler-2002} within GGA, where 
the exchange and correlation functional is parameterized by Perdew, 
Burke and Ernzerhof~\cite{perdew-1996}. It uses the norm-conserving 
pseudopotentials of Troullier and Martins \cite{troullier-1991}. The 
localized basis functions of ``double-$\zeta$ with polarization orbitals'' 
quality (and triple-$\zeta$ for $3d$ functions) have been 
constructed according to the standard scheme of the 
{\sc Siesta} method \cite{junquera-2001}, with the ``Energy Shift'' 
parameter of 0.01~Ry. The treatment of SOI was included as described by 
Fernandez-Seivane \emph{et al.} \cite{fernandez-2006}. 
In the {\sc Siesta} calculations, we used the atomic coordinates
as relaxed by {\sc Vasp}. This allows for a
direct comparison of non-collinear
structures, spin and orbital moments. The standard representation 
of local magnetic moments in {\sc Siesta} is in terms of Mulliken populations
of localized orbitals, that is a quite different definition from that
used in {\sc Vasp}. For the sake of better comparison, we report
in the following the magnetic moments as properties integrated over
atom-centred spheres for both {\sc Vasp} and {\sc Siesta}.       
\begin{table}[t,h]
\caption{\label{tab:coordM13}
Relaxed coordinates ({\AA}) for Jahn-Teller distorted (JT) 
and Mackay transformed (MT) Fe$_{13}$ clusters. See Fig. \ 1 for an 
illustration and labelling of atoms.  
}
{\scriptsize
\begin{tabular}{cdddcddd}
\hline
\hline
\rule[0mm]{0mm}{4mm}
         & \multicolumn{3}{c}{Fe$_{13}$(JT)} &
         & \multicolumn{3}{c}{Fe$_{13}$(MT)} \\
           \cline{2-4}  \cline{6-8}   
Atom Nr. & \multicolumn{1}{c}{~~$x$} & 
           \multicolumn{1}{c}{~~$y$} & 
	   \multicolumn{1}{c}{~~$z$} &
         & \multicolumn{1}{c}{~~$x$} & 
           \multicolumn{1}{c}{~~$y$} & 
	   \multicolumn{1}{c}{~~$z$} \\
         
\hline
1    &  0.6764 & -2.0818 & -1.0306 &  
     &  0.0000 & -1.2917 & -2.0157 \\  
2    & -0.6764 & 2.0818 & 1.0306 &  
     &  0.0000 &  1.2917 &  2.0157 \\  
3    &  0.0000 &  0.0000 & -2.3361 &  
     &  0.0000 &  1.2916 & -2.0157 \\  
4    &  0.0000 & 0.0000 & 2.3361 &  
     &  0.0000 & -1.2916 & 2.0157 \\ 
5    & -0.6764 & -2.0818 & 1.0306 &  
     & -1.2917 & -2.0157 &  0.0000 \\  
6    & 1.7709 & -1.2866 & 1.0306 &  
     & 1.2916 & -2.0157 &  0.0000 \\ 
7    & -1.7709 & 1.2866 & -1.0306 &  
     & -1.2916 &  2.0157 &  0.0000 \\  
8    &  0.6764 &  2.0818 & -1.0306 &  
     &  1.2917 &  2.0157 &  0.0000 \\  
9    & -1.7709 & -1.2866 & -1.0306 &  
     & -2.0157 &  0.0000 & -1.2917 \\  
10   &  2.1890 &   0.0000 &  -1.0306 &  
     &  2.0157 &  0.0000 & -1.2916 \\  
11   & -2.1890 & 0.0000 & 1.0306 & 
     & -2.0157 &  0.0000 &  1.2916 \\  
12   & 1.7709 &  1.2866 & 1.0306 &  
     &  2.0157 &  0.0000 &  1.2917 \\  
13   &  0.0000 &  0.0000 &  0.0000 &  
     &  0.0000 &  0.0000 &  0.0000 \\  
\hline
\hline
\end{tabular}
}
\end{table}

Finally, several test calculations have been performed by a third
method in order to confirm the presented results.
These calculations are documented later in Section IV and have been 
carried out with the all-electron local-orbital 
code ({\sc Fplo} 6.00-24)~\cite{klaus-1999}
in its cluster mode \cite{FPLO} using LSDA \cite{perdew-1992}. 
The valence basis comprised $3s3p3d\; 4s4p4d\; 5s$ states
for $3d$-metals, while $5s5p5d\; 6s6p6d\; 7s$
states were used for Pt. The fully relativistic 
mode is employed here, where {\sc Fplo} solves the four-component 
Dirac-Kohn-Sham equations including spin-orbit coupling (at the 
single electron level) in all orders \cite{Eschrig:2004}. Default
settings were used for the other numerical parameters.

\section{Results for monometallic icosahedral clusters M$_{13}$}
In order to systematically pursue a search towards probable relaxation
patterns from the ideal icosahedron (ICO) structure, we ``drove'' the
structure along two different paths, which are known from previous
studies~\cite{rollmann-2007, rollmann-2004, rollmann-2006, rollmann-th-2007} and are likely to lead to different metastable arrangements.
The structural relaxation of M$_{13}$ clusters results 
in Jahn-Teller (JT) distortion as well as 
partially Mackay transformed (MT) clusters 
(Fe, Co, Ni), which is shown in Fig.\ 1 for the case of Fe$_{13}$. For 
the Fe$_{13}$ cluster, the JT-distorted structure is by 123 meV/cluster, 
the MT cluster by 35 meV/cluster lower in energy compared to the energy 
of the ideal ICO, see also Ref.~\cite{rollmann-2007}. Only 
the Fe cluster exhibits the JT-distortion and the Mackay distortion
as additional local minimum in the energy 
curve~\cite{rollmann-2004, rollmann-2006}, 
whereas Co and Ni show only a tendency for MT.

What we refer to in the following as the Jahn-Teller (JT) type 
distortion~\cite{jahn-1937}
is that which maintains the fivefold symmetry around one of the
ICO axes; it allows a compression or tension of the cluster along
this axis, possibly accompanied by a mutual opposite rotation of two
pentagonal rings pierced by the axis in question. 

The occurrence of the JT distortion originates from (accidental) 
quasi-degeneracy of the highest occupied molecular orbitals so that 
the system may lower its energy by level splitting arising from the 
induced distortion and corresponding redistribution of electrons. 
The partially Mackay distortion is of completely different 
origin since it is connected with the tendency of the magnetic 
Fe clusters to transform to the fcc cuboctahedron (CUBO) and by subsequent 
Bain like transformation to bcc Fe. Whereas for large Fe clusters, 
the JT distortion is no longer observed, we find in the 
simulations that the partially Mackay transformed 
cluster (which were denoted as shell-wise Mackay transformed cluster 
in~\cite{rollmann-2007, rollmann-th-2007}) still corresponds to a 
metastable~\cite{entel-2008} (local energy minimum) state up to cluster sizes 
having 15 closed atomic shells ($n$) and magic atom numbers (N) defined by
\begin{equation*}
N = {\textstyle\frac{1}{3}}(10n^3+15n^2+11n+3).
\end{equation*}

For the Fe$_{13}$ cluster, the JT distortion corresponds to a 
compression of the cluster along the ${\it z}$- axis as depicted 
in Fig.\ 1, whereas we have chosen a different orientation for the MT 
cluster in the same figure (${\it z}$-axis through mid-points of 
two opposite bonds) in order to illustrate its more cubic-like appearance. 
The relaxed atomic coordinates for the MT and JT 13-atom Fe clusters 
are listed in Table I. For the JT-distorted 
Fe$_{13}$ cluster, 10 of the peripheral atoms lie at a distance of 2.42~{\AA}
from centre, the other two atoms show an inward relaxation towards 
the centre and hence, remain at a distance of 2.34~{\AA} from centre.
A clear picture of the relaxation is shown in the left panel of Fig.\ 1, 
where the arrows associated with atoms on the top and bottom of the 
box face oppositely. The structural relaxation of the Fe$_{13}$ cluster 
shown in the right panel in Fig.\ 1 corresponding to the partial Mackay 
transformation is the onset of the transformation from an ICO 
to CUBO, where the consecutive triangular facets of the 
ICO change to square facets of the CUBO, which 
can be described by a parameter ${\it s}$ defined as the  square 
of the ratio of the stretched to the unstretched 
edges being 1 and 2 for the ICO and CUBO, respectively. 
See~\cite{mackay-1962} for the original explanation. It is well 
observed for multishell ICO-like Fe 
clusters~\cite{rollmann-2007, rollmann-th-2007}. In our
calculations, the relaxed Fe$_{13}$ has ${\it s}$ = 1.07; Co$_{13}$ and 
Ni$_{13}$ have ${\it s}$ ${\approx}$ 1, resulting in an almost ideal 
icosahedral like structure.  

The peripheral bond lengths for the MT clusters (see Table II) can 
be arranged in groups: 6 of the 30 bonds being parallel
to the cartesian axes (${\it x}$: 5-6, 7-8; ${\it y}$: 1-3, 4-2; ${\it z}$: 
9-11, 10-12; see Fig.\ 1 for the labeling of the atoms), the remaining 
24 bond lengths are also listed in Table II. Remarkably, the distances 
of the 12 peripheral atoms to the central one are all 
identical. 

\begin{table}
\caption{\label{tab:table2}The bond lengths in \AA~for relaxed 
Mackay transformed M$_{13}$ clusters.}
\begin{ruledtabular}
{\scriptsize
\begin{tabular}{lccr}
Bonds &Fe$_{13}$&Co$_{13}$&Ni$_{13}$\\
\hline
Centre-shell & 2.394 & 2.33 & 2.321 \\
24 $\times$ & 2.501 & 2.451 & 2.441 \\
6 $\times$ & 2.583 & 2.447 & 2.439 \\
\end{tabular}
}
\end{ruledtabular}
\end{table}
\begin{table}
\caption{
\label{tab:table3}Spin and orbital moments of the relaxed
Fe$_{13}$ (MT) and Co$_{13}$ clusters as calculated by {\sc Vasp} 
and {\sc Siesta}, for the 
initial $[001]$ setting of spin direction. The $x$ and $y$-components of 
the spin moment are $<$ 0.001, hence not shown here. For {\sc Siesta} 
calculations, the values of spin moments are given in terms of 
Mulliken populations, the magnetic moments are in units of 
$\mu_{B}$/atom.
}

\begin{ruledtabular}
{\scriptsize
\begin{tabular}{ccccccccc}
&\multicolumn{3}{c}{Vasp}&\multicolumn{5}{c}{Siesta}\\
\hline
&\multicolumn{7}{c}{Fe$_{13}$}\\
\hline
Atom No.&$\mathrm{L}_x$&$\mathrm{L}_y$&$\mathrm{L}_z$&$\mathrm{S}_z$&$\mathrm{L}_x$&$\mathrm{L}_y$&$\mathrm{L}_z$&$\mathrm{S}_z$\\ \hline 
 1& 0.00 &-0.02 & 0.11 & 3.06 & 0.00 &-0.01 & 0.09 & 3.43 \\
 2& 0.00 & 0.02 & 0.12 & 3.06 & 0.00 & 0.01 & 0.09 & 3.43 \\
 3&-0.01 & 0.00 & 0.08 & 3.08 &-0.01 & 0.00 & 0.08 & 3.43 \\
 4& 0.01 & 0.00 & 0.08 & 3.08 & 0.01 & 0.00 & 0.08 & 3.44 \\
 5& 0.00 & 0.00 & 0.07 & 3.10 & 0.00 & 0.00 & 0.06 & 3.44 \\
 6& 0.00 & 0.00 & 0.07 & 3.10 & 0.00 & 0.00 & 0.06 & 3.44 \\
 7& 0.00 & 0.00 & 0.07 & 3.10 & 0.00 & 0.00 & 0.06 & 3.44 \\
 8& 0.00 & 0.00 & 0.07 & 3.10 & 0.00 & 0.00 & 0.06 & 3.44 \\
 9& 0.01 & 0.00 & 0.08 & 3.08 & 0.01 & 0.00 & 0.08 & 3.44 \\
10&-0.01 & 0.00 & 0.08 & 3.08 &-0.01 & 0.00 & 0.08 & 3.44 \\
11& 0.00 & 0.02 & 0.11 & 3.05 & 0.00 & 0.01 & 0.09 & 3.43 \\
12& 0.00 &-0.02 & 0.11 & 3.06 & 0.00 &-0.01 & 0.09 & 3.43 \\
13& 0.00 & 0.00 & 0.05 & 2.70 & 0.00 & 0.00 & 0.03 & 2.75 \\
\hline
&\multicolumn{7}{c}{Co$_{13}$}\\
 \hline
 1& 0.00 & 0.03 & 0.09 & 2.10 & 0.00 & 0.01 & 0.08 & 2.42 \\
 2& 0.00 &-0.03 & 0.09 & 2.10 & 0.00 &-0.00 & 0.08 & 2.41 \\
 3& 0.02 & 0.00 & 0.14 & 2.10 & 0.02 & 0.00 & 0.10 & 2.42 \\
 4&-0.02 & 0.00 & 0.14 & 2.10 &-0.02 & 0.00 & 0.10 & 2.42 \\
 5& 0.00 & 0.00 & 0.17 & 2.10 &-0.00 &-0.00 & 0.13 & 2.41 \\
 6& 0.00 & 0.00 & 0.17 & 2.10 & 0.00 &-0.00 & 0.13 & 2.42 \\
 7& 0.00 & 0.00 & 0.17 & 2.10 & 0.00 &-0.00 & 0.13 & 2.42 \\
 8& 0.00 & 0.00 & 0.17 & 2.10 & 0.00 &-0.00 & 0.13 & 2.41 \\
 9&-0.02 & 0.00 & 0.14 & 2.10 &-0.02 & 0.00 & 0.10 & 2.42 \\
10& 0.02 & 0.00 & 0.14 & 2.10 & 0.02 & 0.00 & 0.10 & 2.42 \\
11& 0.00 &-0.03 & 0.09 & 2.10 & 0.00 &-0.00 & 0.08 & 2.41 \\
12& 0.00 & 0.03 & 0.09 & 2.10 & 0.00 & 0.01 & 0.08 & 2.42 \\
13& 0.00 & 0.00 & 0.05 & 1.91 & 0.00 & 0.00 & 0.04 & 1.96 \\
\end{tabular}
}
\end{ruledtabular}
\end{table}
\begin{table}
\caption{\label{tab:table2}Cluster averaged values for 
${\langle\rm{L}\rangle}$ = $\frac{1}{13}$$\sum_{i=1}^{13}{|\mathbf{L}_i|}$ 
and ${\langle\rm{S}\rangle}$ = $\frac{1}{13}$${\sum_{i=1}^{13}}{|\mathbf{S}_i|}$ of MT M$_{13}$ clusters are shown in ${\mu_B}$/atom,
compared to bulk values for bcc Fe, fcc Co and Ni as obtained 
from our calculations.}
\begin{ruledtabular}
{\scriptsize
\begin{tabular}{lcccccr}
Cluster &${\langle{\rm L}\rangle}$&${\langle{\rm S}\rangle}$&${\langle|{\mathbf L}|\rangle}_{\mathrm {bulk}}$&${\langle|{\mathbf S}|\rangle}_{\mathrm {bulk}}$\\
\hline
Fe$_{13}$      & 0.084 & 3.04 & 0.06 & 2.25 \\
Co$_{13}$      & 0.125 & 2.08 & 0.08 & 1.67 \\
Ni$_{13}$      & 0.070 & 0.66 & 0.05 & 0.65 \\
\end{tabular}
}
\end{ruledtabular}
\end{table}
\begin{table}
\caption{\label{tab:table2}The values of prefactors in the Heisenberg anisotropy term in Eq. (2) for the relaxed M$_{13}$ clusters.}
\begin{ruledtabular}
{\scriptsize
\begin{tabular}{lccr}
Cluster &D${_2}$ (meV)&D${_4}$(meV)&D${_6}$$(\mu eV)$\\
\hline
Fe$_{13}$ (ICO) (46 $\mu$$_B$) & -- & -- & -41.1 \\
Fe$_{13}$ (JT) (44 $\mu$$_B$) & 15.2 & -- & -- \\
Fe$_{13}$ (MT) (44 $\mu$$_B$) & -- & -11.5 & -41.1 \\
Co$_{13}$ (31 $\mu$$_B$)     & -- & -21.7 & -25.1 \\
Ni$_{13}$ (8 $\mu$$_B$)     & -- & 7.1 & 7.3 \\
\end{tabular}
}
\end{ruledtabular}
\end{table}
The relaxed MT coordinates obtained with {\sc Vasp} have been used in the 
calculation by {\sc Siesta}, without further relaxation, in order 
to compare the resulting values of spin and orbital moments, and 
the non-collinearity. Table III compares the calculated results by 
the two methods for Fe$_{13}$(MT) and Co$_{13}$ (MT) clusters. 
For both methods, the spin moments remain parallel ({$x$}- 
and {$y$}-components of the spin vectors are, at most, 0.001 $\mu$$_B$, 
and are not shown in Table III). The orbital moments for MT distorted 
Fe$_{13}$ obtained by both programs are directed towards the  
$z-$direction in the lower part of the cluster and are 
deviated ``outwards'' from the $z-$direction in the upper part, while
for Co$_{13}$, they show opposite behavior, i.e. the orbital moments 
deviate away from the $z-$axis in the lower part of 
the cluster and they are directed towards the $z-$axis in the upper part.
but we note that these deviations are not significantly above 
the numerical noise level. This becomes evident if one compares data for
those atoms that are mutually equivalent (apart from
rotations around the z-axis) by symmetry: 1-4, 5-8, 9-12. While the
symmetry requirement is (almost) obeyed by the atoms 5-8, deviations
of a few 1/100 $\mu_B$ are found between the atoms 1-4 and also
between the atoms 9-12.
 
When comparing the numerical results for spin and orbital magnetic moments from
these two different calculational methods, one must take into acount
the difference in their definition. In {\sc Vasp}, the properties (spin and
orbital moments) are extracted as projection onto an atomic sphere
(see Section~\ref{sec:methods}). The ``standard'' {\sc Vasp} value
of atomic sphere radius for Co does in fact correspond to 
slightly overlapping spheres in our Co$_{13}$ cluster.
The {\sc Siesta} output results are reported in terms of 
decomposition over projection onto
localized, but spatially extended, numerical
orbitals, known as Mulliken population analysis.
It is known that the local magnetic moments as well as atomic
charges in heterogeneous systems do often come out very different, when
estimated according to these two different schemes. In order to
illustrate this effect, we give in the last column 
of Table III the values of spin moment, extracted from 
the {\sc Siesta} results. The fluctuations of these integrated 
values over apparently equivalent atoms are caused by the sparseness 
of the spatial grid with the step 0.078~{\AA}, on which the summation 
of spin density has been done.

The ${\langle \rm{L} \rangle}$ and ${\langle \rm{S} \rangle}$ of 
Fe$_{13}$, Co$_{13}$ and Ni$_{13}$ clusters are 
compared with respect to bulk in Table IV. The orbital and spin moments
for bulk systems are calculated with the equlibrium lattice constants of 
2.832~{\AA} (bcc Fe), 3.52~{\AA} (fcc Co) and 3.523~{\AA} (fcc Ni).
The orbital and spin magnetic moments of the elemental Fe, Co and Ni 
clusters are larger than the related bulk values.
For example, in Ref~\cite{guirado2003} it was shown that 
in Ni clusters of up to 13 atoms, the average 
orbital moment ${\langle {\mathrm L} \rangle}$ per atom is found to be 4 to 8 
times larger with respect to L$_{\rm bulk}$;  
for larger cluster sizes, ${\langle {\mathrm L} \rangle}$ was shown 
to approach the bulk value.
It turns out from our calculation that for Ni$_{13}$ clusters, 
${\langle {\mathrm L} \rangle}$ agrees quite well with the earlier 
report~\cite{guirado2003}, with orbital moment having larger 
magnitude than the bulk value. For Fe$_{13}$ and Co$_{13}$, the 
${\langle \rm{L} \rangle}$ is also relatively large with respect to 
the bulk systems.
In addition, the average spin moments ${\langle \rm{S} \rangle}$ 
for these clusters are also increased with respect to the bulk systems.

In addition, the magnetic anisotropy energy for M$_{13}$
clusters was calculated. Bulk $3d$ metals show a tiny
(cubic Fe, Co and Ni) MAE of $\sim$1~$\mu$eV.
The tiny value in cubic systems is due to the high
symmetry and it is expected that an ideal ICO
exhibits a similarly small MAE. If the symmetry is broken,
e.g., by tetragonal distortion of a cubic system,
the MAE increases considerably (see for instance Ref. 
\onlinecite{Gruner2008}).

Figure 2 shows the ideal icosahedral cluster and the definition of 
the ($x,z$)-plane with angle $\theta$ for the magnetization 
directions used in our MAE calculations. 
The magnetic anisotropy energy is computed from the total energy 
as $\Delta E= E^{tot}(\theta)-E^{tot}({\theta=0})$, 
$E^{tot}(\theta)$ being the $\theta$-dependent total energy.
 The $\theta$-dependent energy obtained from total energy
calculations is compared with the anisotropy term of the extended
Heisenberg model, defined as

\begin{equation}
H = -\sum_{i \neq j}J_{ij}{\bf S}_{i}{\bf S}_{j}+H_{\mathrm{ani}}+H_{\mathrm{dipolar}},
\end{equation}
where
\begin{equation}
H_{\mathrm{ani}} = -\sum_{n}\sum_{i}^{N}D_{n}({\bf e}_{i} \cdot {\bf S}_{i})^n. 
\end{equation}

The spin-spin and dipolar interactions are neglected, while comparing 
{\it ab-initio} total energy differences with the Heisenberg model.

In Eq. (2), $n$ is the order of anisotropy (an even integer), 
${\bf e}_i$ are the unit vectors of the atoms along the radial 
directions, {\bf S}$_i$ their spin moments and $D$ is the 
anisotropy constant (of N\'{e}el-like model), which is 
assumed to be $\theta$-independent. The second and fourth order 
contributions are constant from symmetry considerations of a ideal 
ICO. The anisotropy consists then mostly of a sixth order 
contribution, 
assuming that still higher orders are negligibly small. 
Figure 3 shows a comparison of energy differences of an ideal Fe$_{13}$  
ICO obtained from {\it ab-initio} calculations to that 
of the Heisenberg model using a least-mean square fit to 
the {\it ab-initio} data. It is a delicate task 
to correctly determine energy contributions in the
order of a few $\mu$eV/atom or below. To obtain meaningful results
presupposes that the charge density is extremely well converged and
still consistent with the symmetry of the system. 
A distortion of the symmetry might occur due to numerical fluctuations
(which can to a limited extent be controlled by, e.g., the choice of
energy cutoff and Fourier-grid) and to perturbations by the setup of the
model system itself, e.g., due to the electrostatic interaction between
periodic images of the supercell. 
Suitable  settings can be found from the corresponding contributions to 
the forces, which depend on the ionic positions and the charge distribution
as well. Consequently, the magnitude of the forces acting on all
equivalent atoms should be nearly equal, which in our case, we have 
taken care that forces are not larger than 10$^{-6}$ eV/{\AA}. 
It is shown from model calculations of fcc Co clusters that 
the $\theta$-dependence of magnetic anisotropy may also differ 
with the polar angle $\phi$ \cite{yane-2009}, which is not discussed here.
\\ 

Along with the MAE for a ideal cluster shown in Fig.\ 3, the MAE of relaxed 
M$_{13}$ clusters is calculated: The variation of energy difference  
with respect to $\theta$ for relaxed clusters are shown in Fig.\ 4. 
The $\theta$-dependent energy difference for the JT-distorted Fe$_{13}$
cluster (the top panel of Fig.\ 4) gives rise to a 
significant D$_2$ contribution of $\sim$ 15.2 meV 
for the anisotropy. However, approximately only a D$_4$ and D$_6$ 
contribution 
arises for the anisotropy in case of
MT Fe$_{13}$ cluster. The values are depicted in Table V. The MAE for the 
JT-distorted Fe cluster is calculated to be 322 $\mu$eV/atom, which 
is approximately five times larger with respect to the MT 
cluster (MAE$_{MT}$ = 60 $\mu$eV/atom). The reason behind the large MAE 
for the JT-distorted cluster relative to the MT cluster may be 
attributed to the breaking of symmetry in the former case. 
   
The $\theta$-dependent energy difference for the relaxed MT Fe$_{13}$ 
cluster shows qualitatively good resemblance with that of the ideal cluster
(see Fig.\ 3), however, the quantitative MAE value for ideal 
Fe$_{13}$ ICO is found 
to be approximately 1 $\mu$eV/atom, while for the MT Fe$_{13}$ cluster, it is 
approximately 60 times larger than for the ideal cluster. However, 
for relaxed geometries of Co$_{13}$ and Ni$_{13}$, the MAE value 
is calculated to be 5 and 0.64 $\mu$eV/atom, respectively. Our 
calculations suggest a larger value of MAE per atom for Fe$_{13}$ 
and Co$_{13}$ compared to the bulk, which is 1.4 $\mu$eV/atom (bcc Fe), 
1.3 $\mu$eV/atom (fcc Co) \cite{halilov-1998}. For Ni$_{13}$, this is 
lower with respect to bulk fcc Ni (2.7 $\mu$eV/atom). The small anisotropy 
for the relaxed Ni$_{13}$ cluster relative to bulk Ni is 
due to the high symmetry of the cluster, Ni$_{13}$ having a nearly 
ideal icosahedral shape. The change of sign in the 
energy vs. $\theta$-curves (shown in Fig.\ 4) for the relaxed 
M$_{13}$ clusters can be explained through their way of structural 
relaxation. Namely, the MT relaxation pattern for the 
Co$_{13}$ cluster is just opposite to that of Fe$_{13}$: The 
Co "dimers" oriented along the edges of the cube
in the right panel of Fig.\ 1 make the shortest side in each 
Co$_3$ face, whereas the corresponding Fe-Fe bonds are the longest 
in the Fe$_3$ faces of the MT Fe$_{13}$ cluster, see Table II. 
By comparing {\it ab-initio} results with the Heisenberg anisotropy
term, the D$_4$ of M$_{13}$ can be fitted, which are listed in Table V. 
It is observed that D$_4$ makes a significant contribution 
to the magnetic anisotropy of the relaxed clusters having 
cubic-like distortions.

It would be fascinating to see how spin and orbital moment as well as 
the MAE change, if we consider more ``asymmetric'' clusters. To this
we have added a varying number of Pt atoms on the top of M$_{13}$ clusters.   

\begin{figure*}
\includegraphics[scale=0.15]{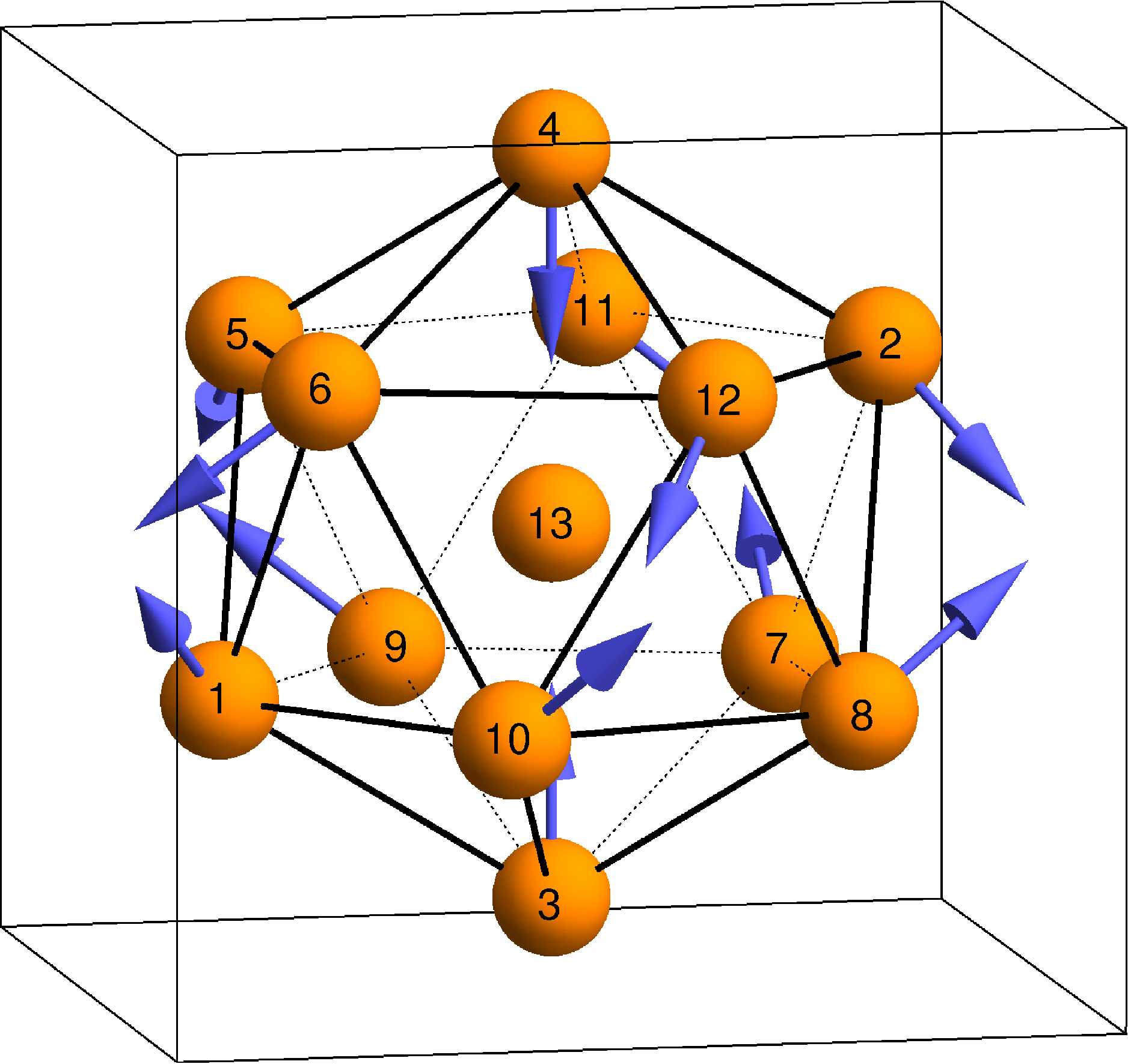}
\hspace{0.7cm}
\includegraphics[scale=0.14]{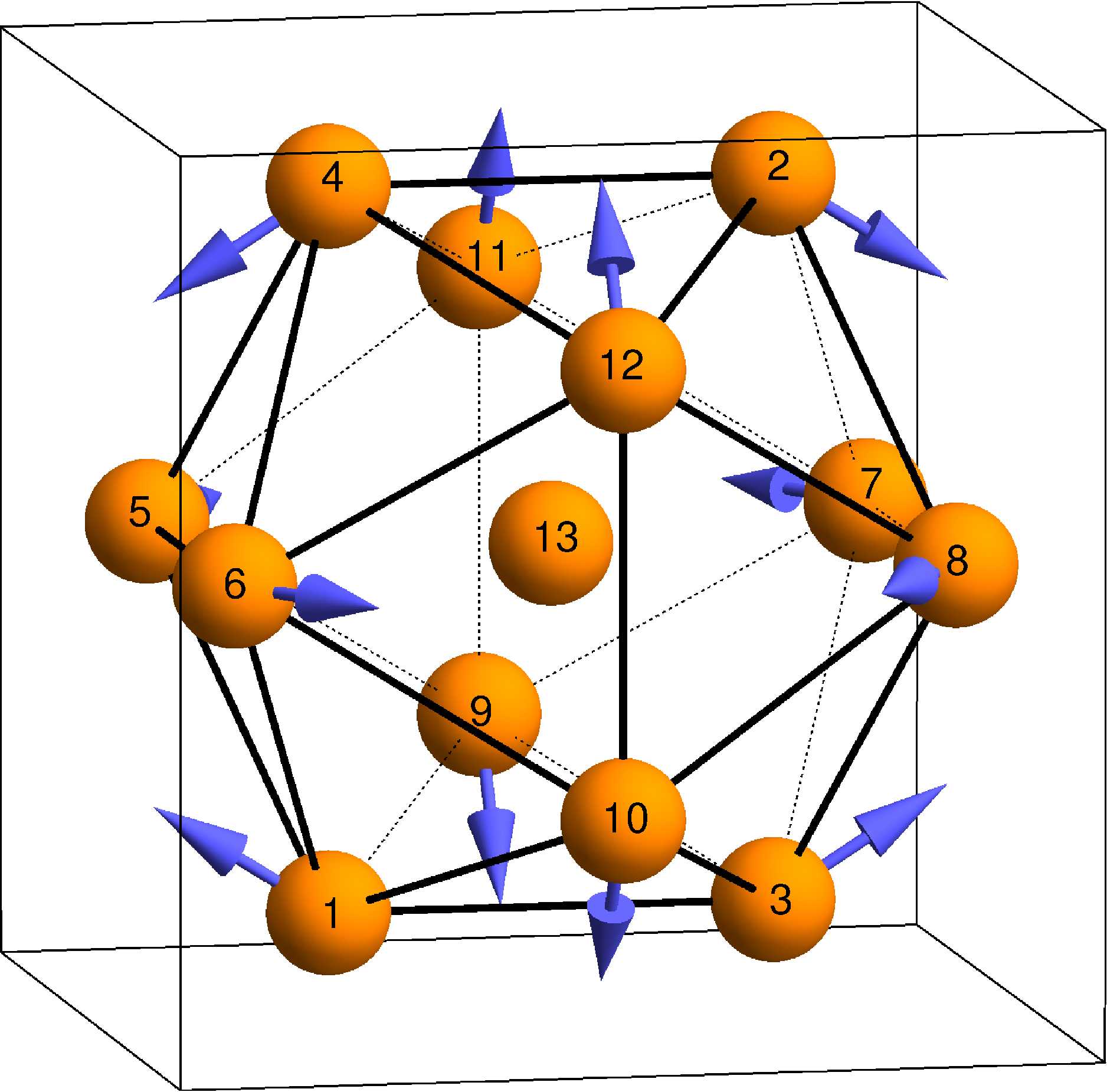}
\caption{\label{fig:epsart} The JT-distorted (left) and the MT (right) 
Fe$_{13}$ cluster. Arrows indicate the direction of relative shift of 
atoms with respect to the ideal positions. For the JT-distorted 
and MT Fe$_{13}$ cluster, the displacements marked by the arrows 
have been scaled up by a factor of 20 and 30, respectively. The box 
is only guide to the eyes. (The actual simulation 
box size is 15$^3$~\AA$^3$.)   
}
\end{figure*}
  
\begin{figure}
\includegraphics[scale=0.12]{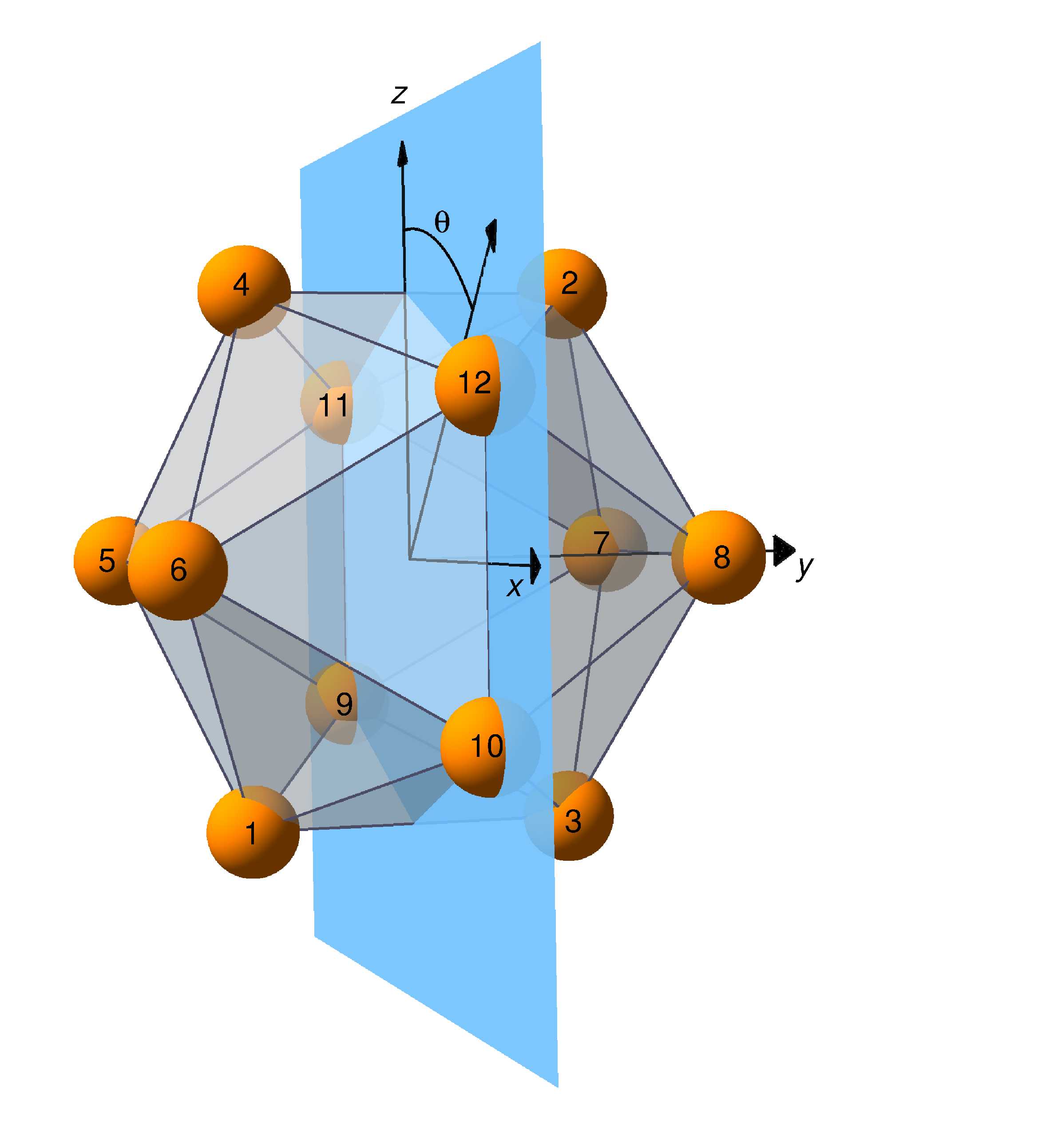}
\caption{\label{fig:epsart} The ideal icosahedral cluster 
showing the $x$-$z$ plane in which the angle $\theta$ is varied in 
the MAE calculations for the partially Mackay transformed clusters.}
\end{figure}

\begin{figure}
\includegraphics[scale=0.50]{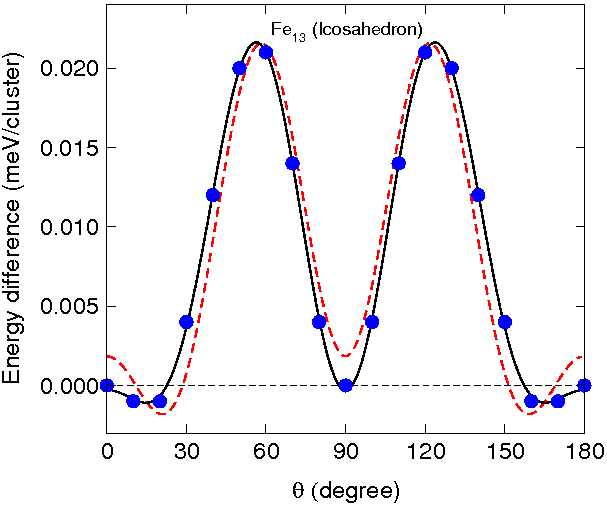}
\caption{\label{fig:epsart} The $\theta$-dependent  
energy for an ideal Fe$_{13}$ icosahedral cluster showing the 
{\it ab-initio} data (filled circles and solid line as guide to 
the eyes) compared with 
the Heisenberg model (dashed curve). For {\sc Vasp} calculations, 
the energy difference is in units of meV per cluster and is 
defined as $\Delta E= E-E_{\theta=0}$.
}
\end{figure}

\begin{figure}
\includegraphics[scale=0.65]{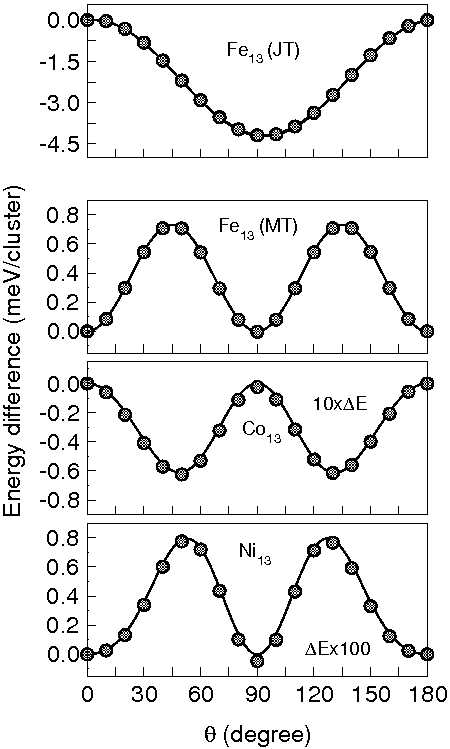}
\caption{
\label{fig:epsart}The plots from top to bottom 
show the energy differences in meV/cluster vs. $\theta$ of relaxed Fe$_{13}$ 
(for two types of relaxations - JT and 
Mackay distortion), Co$_{13}$ and Ni$_{13}$ clusters. In case of 
JT-distorted Fe$_{13}$: $\theta$ varies with respect to the $z$-axis in 
the plane passing through atom 4, middle of the bond 
between atoms 6-12, atom 10 and atom 3 (shown in the left side of Fig.\ 1), 
whereas for the MT cluster, $\theta$ varies through all atoms as depicted 
in Fig.\ 1. Energy difference is ($\Delta E= E-E_{\theta=0}$). 
The energy difference for the Co$_{13}$ cluster is multiplied by 
a factor 10 and the energy difference for the
Ni$_{13}$ cluster is multiplied by a factor 100.
} 
\end{figure}

\begin{figure}
\includegraphics[scale=0.3]{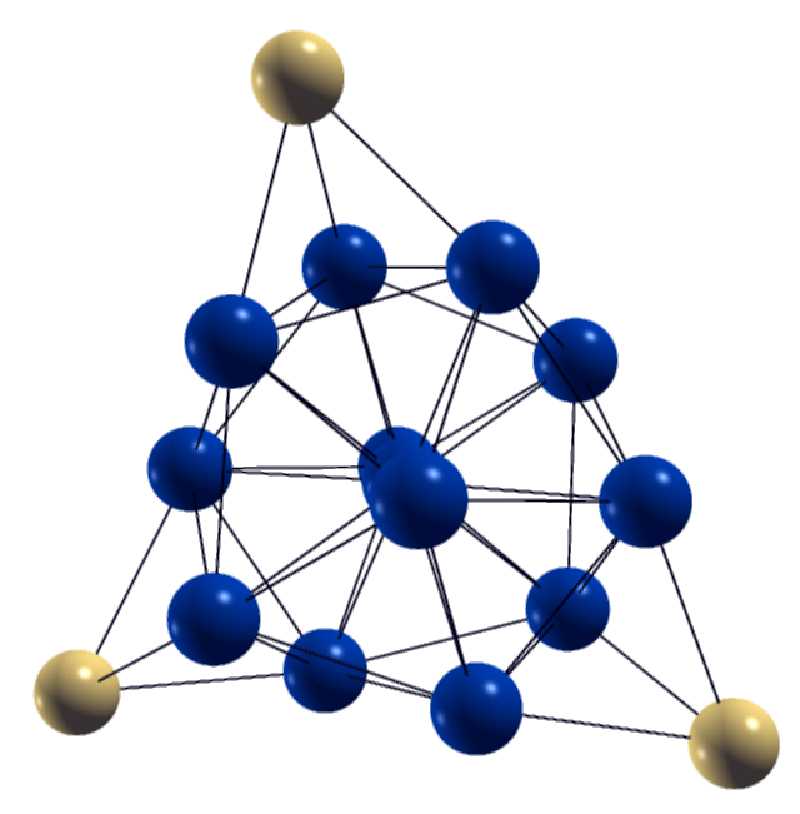}
\includegraphics[scale=0.3]{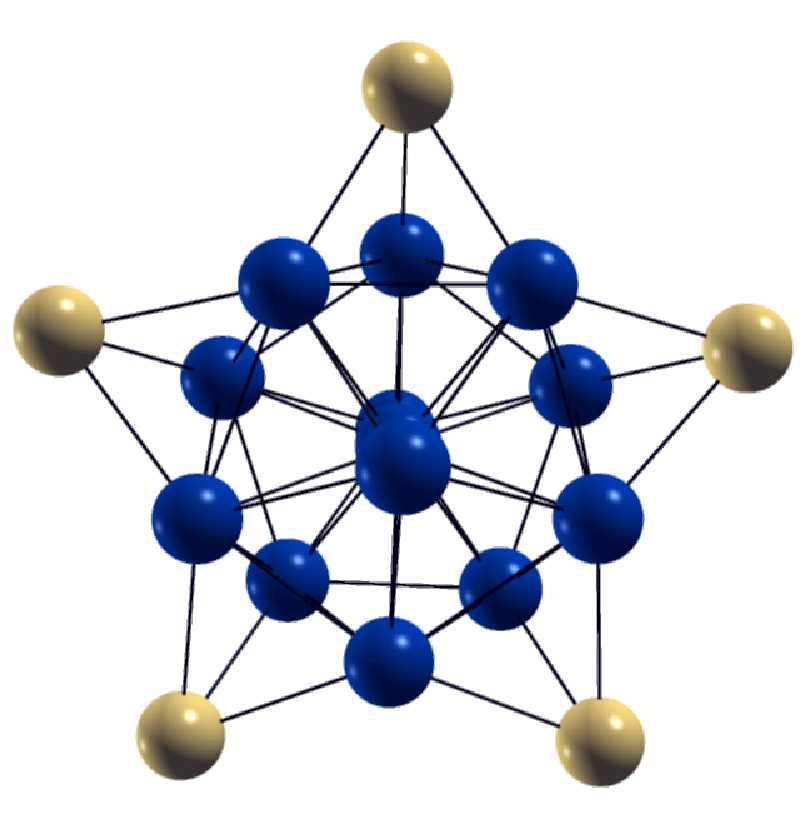}
\caption{\label{fig:epsart}The relaxed structure of 
Co$_{13}$Pt$_3$ (left) and Co$_{13}$Pt$_5$ (right) shows the 
arrangement of Pt atoms on the Co$_{13}$ cluster, dark (blue) spheres: 
Co atoms and light (yellow) spheres: Pt atoms. In each case all Pt atoms 
lie in the same plane corresponding to the lowest energy structures 
found so far.}  
\end{figure}

\begin{figure}
\includegraphics[scale=0.3] {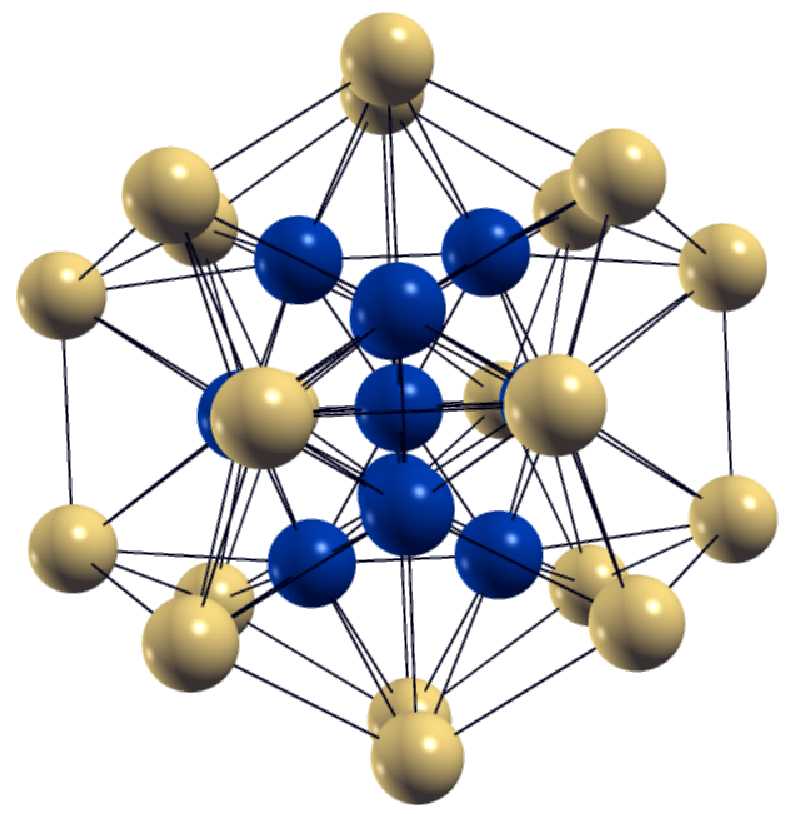}
\includegraphics[scale=0.3] {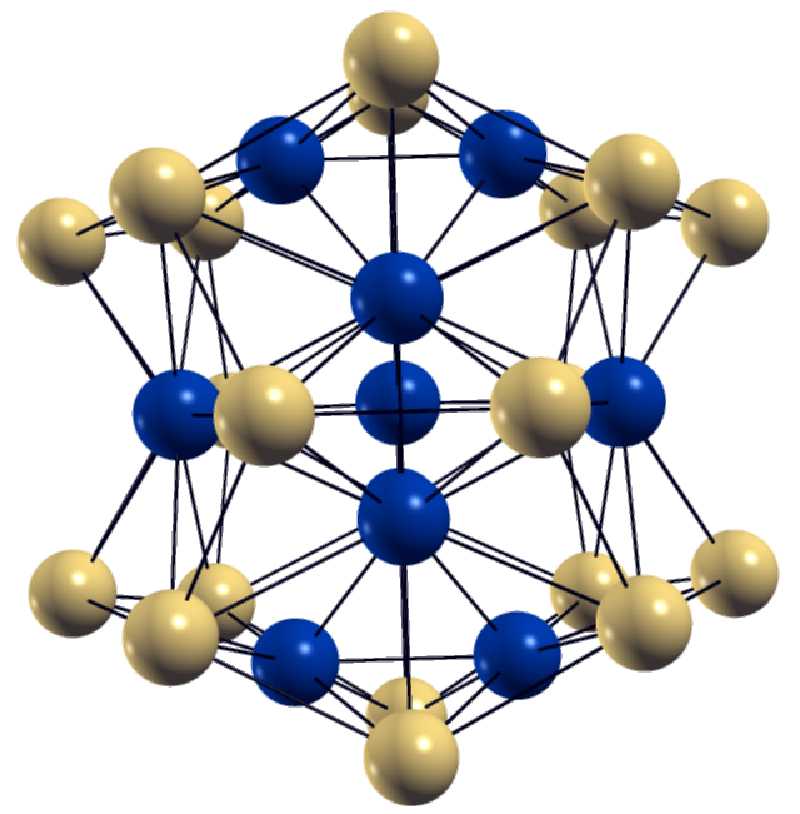}
\caption{\label{fig:epsart}The initial and final strucutres for the
Ni$_{13}$Pt$_{20}$ cluster are shown in the left and right panel, 
respectively. Ni and Pt atoms are represented by dark (blue) 
and light (yellow) spheres, respectively. The final structure, 
corresponding to the lowest energy structure found so 
far, demonstrates the importance of atomic relaxations.}
\end{figure}

\begin{figure}
\includegraphics[scale=0.60]{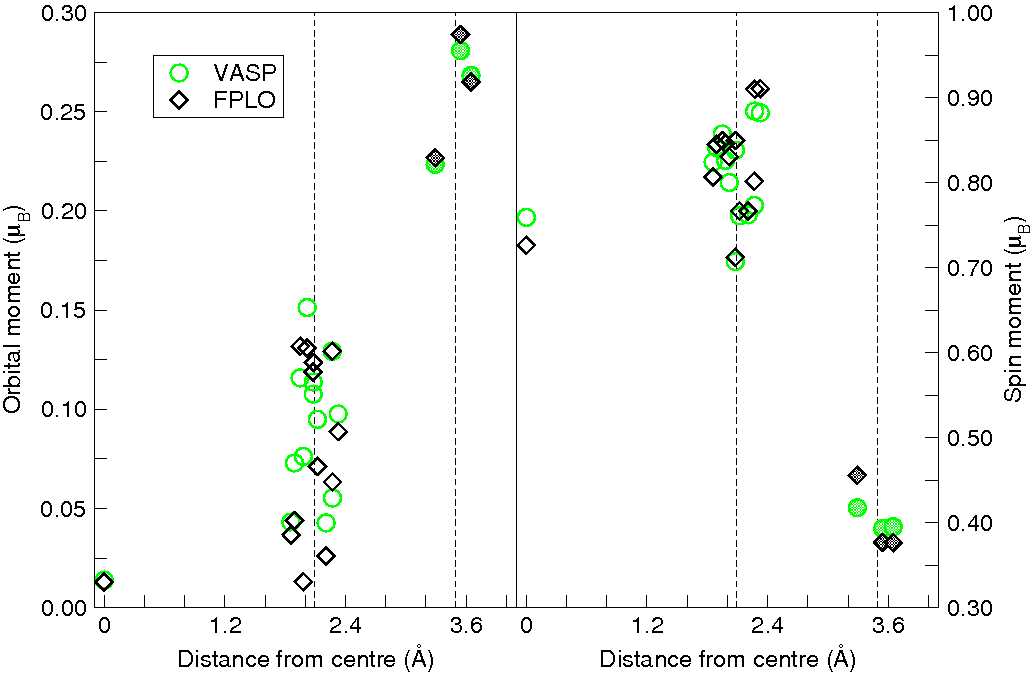}
\caption{\label{fig:epsart}The variation of the individual 
atomic ${|\mathbf{L}_i|}$ (left panel) and ${|\mathbf{S}_i|}$ 
(right panel) with respect to the distance from centre for Ni$_{13}$Pt$_{3}$ 
cluster. $\bigcirc$: results by {\sc Vasp} and 
$\Diamond$: results by {\sc Fplo}. The open and filled 
symbols (for both colors) show the moments on Ni and Pt atoms, respectively.} 
\end{figure}

\begin{figure}
\includegraphics[scale=0.70]{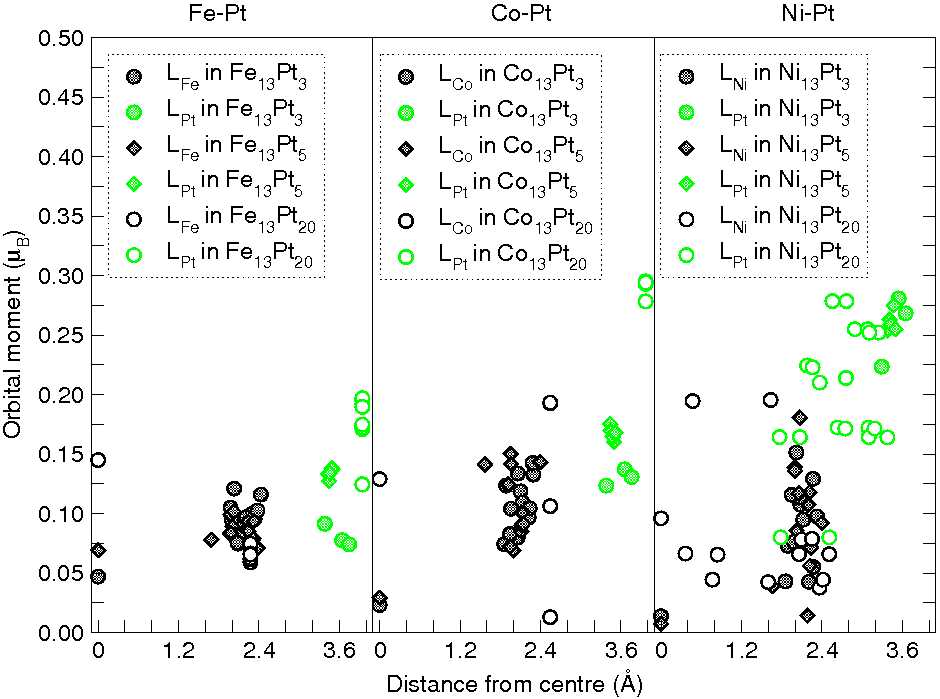} 

\caption{\label{fig:epsart}The variation of the individual atomic orbital moments with respect to distance from centre for Fe$_{13}$Pt$_n$ (left), 
Co$_{13}$Pt$_n$ (middle) and Ni$_{13}$Pt$_n$ (right) clusters. The 
light (green) and dark (black) symbols for all cases represent the ${|\mathbf{L}_i|}$ 
on Pt atoms and  M$_{13}$ clusters, respectively.
}
\end{figure}

\begin{figure}
\includegraphics[scale=0.70]{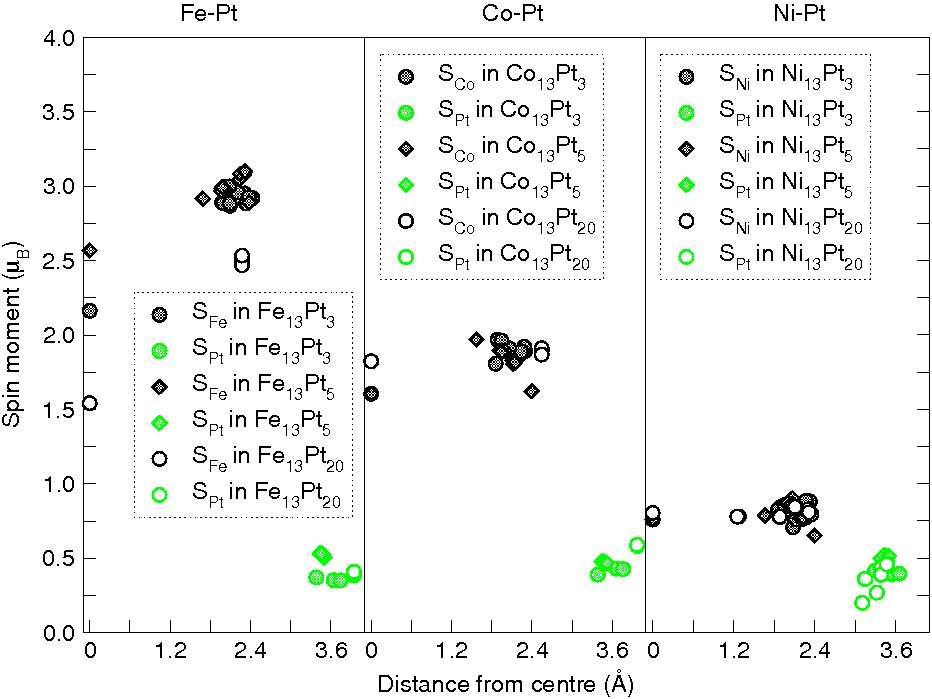}
\caption{\label{fig:epsart}The plots in the left, middle and right 
panels show the variation of $|\mathbf{S}_i|$ with respect to the 
distance from centre for Fe$_{13}$Pt$_{n}$, Co$_{13}$Pt$_{n}$ and 
Ni$_{13}$Pt$_{n}$ clusters, respectively. The same symbols are used here
as in Fig.\ 8.} 
\end{figure}

\section{Results for capped clusters M$_{13}$P\lowercase{t}$_n$}
\label{ref:M13Pt}
Obviously, binary and core-shell transition metal clusters 
show a yet larger diversity
compared to the elemental systems. For example, the intermixing of $3d$ 
elements with $4d$ or $5d$ elements results in a large magnetic 
moment of the binary systems~\cite{navia-2004}.
Both in experiment and in calculations, it has been 
observed that FePt and Co clusters  
show enhanced spin moments and orbital moments 
with respect to corresponding bulk 
values~\cite{boyen-2004, ebert-2006}. Hence, 
it would be interesting to study the change in magnetic properties 
including MAE of M$_{13}$ clusters capping with Pt atoms, which is described in the following.

We considered three high-symmetry positions to cap 
the M$_{13}$ clusters by a single Pt atom and found
that a Pt position above the centre of a facet is most favorable 
in all three cases.
In the following, we used this finding as a guideline for initial geometries
of M$_{13}$Pt$_n$ clusters (n = 3, 4, 5, 20): In all cases, the Pt atoms were
initially placed above the facet centres at a distance found in the single-Pt
capping case. After relaxation, optimised geometries were obtained as 
shown in Fig. 5 for Co$_{13}$Pt$_3$ (left) and Co$_{13}$Pt$_5$ (right) and
in Fig. 6 (right side) for Ni$_{13}$Pt$_{20}$.
For n = 20, the initial geometries form a core-shell morphology.

Since atom projected quantities like spin and orbital moments depend on the
specific code, we compared for the particular case of Ni$_{13}$Pt$_3$ related
data obtained by {\sc Vasp} and {\sc Fplo}. The structure optimisation 
was carried out by VASP and the same geometry was used to evaluate the 
magnetic moments by both codes.

In {\sc Fplo}, the magnetic moments are calculated through Mulliken 
population analysis. Figure 7 shows the absolute value of orbital moment 
per atom ${|\mathbf{L}_i|}$ (left) and the absolute value of spin moment per 
atom ${|\mathbf{S}_i|}$ (right) as a function of distance of each 
atom from the cluster centre for Ni$_{13}$Pt$_3$. 
It is obvious that both codes give results which are in good 
agreement with each other. Therefore, the calculations which are 
discussed in the following are done with {\sc Vasp} only. 

In Figs. 8 and 9, the variation of ${|\mathbf{L}_i|}$ and of 
${|\mathbf{S}_i|}$ with respect to 
the distance from centre of M$_{13}$Pt$_n$ clusters are shown, respectively 
(the symbols are kept consistent for both figures and the centre 
atom is placed at zero). Figure 8 nicely shows that the orbital 
moments of such few-atom
systems in general depend very sensitively on the particular chemical
composition and geometry. The resulting electronic structure can be
very individual (e.g., the nature of the highest occupied level depends
on the electron number and the spin moment) and is hard or impossible
to predict without a detailed calculation.
It is also found that (see Fig.\ 8) ${|\mathbf{L}_i|}$ 
for M$_{13}$ clusters approach towards the corresponding 
bulk values of bcc Fe, fcc Co and Ni reported in Table IV, with 
increasing the number of Pt atoms. The orbital moment on the centre 
Fe atom for Fe$_{13}$Pt$_n$ cluster increases 
with the number of Pt atoms, whereas for Co$_{13}$Pt$_n$ and Ni$_{13}$Pt$_n$, 
this trend is weak. The orbital moment on Pt atoms is found
to be very sensitive with respect to the core atomic species.
For example, the ${|\mathbf{L}_i|}$ on Pt atoms for Ni$_{13}$Pt$_n$ clusters 
are enhanced relative to that of Fe$_{13}$Pt$_n$ and Co$_{13}$Pt$_n$. 

In Fig.\ 9 the variation of spin moments for the capped clusters 
are shown, which suggests that with increasing Pt atoms, the 
${|\mathbf{S}_i|}$ for M$_{13}$ clusters always remain larger 
with respect to the corresponding bulk 3$d$-spin 
moments values. Unlike the trend in orbital moment, 
the spin moment on Pt atoms is not much affected by its core 
species. 
In the following, a few quantitative
statements are made for each of the capped clusters.

\subsection{Fe$_{13}$Pt$_n$ clusters}

The values of average orbital moment and average spin moment for 
each atomic species along with the total orbital moment 
 $ \rm{L_{tot}}$ and total spin 
moment $\rm{S_{tot}}$ for the 
capped clusters are defined and reported in Table VI. Our observation 
suggests that adding more Pt atoms on Fe$_{13}$ cluster, the $\rm{L_{tot}}$ 
for Fe$_{13}$Pt$_n$ and $\rm{S_{tot}}$ for Fe$_{13}$Pt$_3$ and 
Fe$_{13}$Pt$_5$ increase considerably. However, there is a decrease of 
$\rm{S_{tot}}$ for Fe$_{13}$Pt$_{20}$ because of the  
antiferromagnetic like alignment of spin of centre Fe atom with 
respect to the sorrounding atoms in this cluster (for all other 
cappings, the Fe$_{13}$ cluster is ferromagnetic). Capping with 
three and five Pt atoms enhances the atomic orbital moment on Fe$_{13}$ 
by $\sim$ 50 \% and 33 \%, respectively, relative to 
the bulk value of bcc Fe, i.e. 0.06 ${\mu}_B$/atom. 
It does not much change the Fe moment in comparison
with the bare Fe$_{13}$ cluster, however. The average orbital 
moment ${\langle\rm{L_M}\rangle}$ 
on Fe$_{13}$ decreases monotonically with increasing number of Pt atoms.  

\subsection{Co$_{13}$Pt$_n$ clusters}
A ferromagnetic ground state is found to be stable for all compositions of 
Co$_{13}$Pt$_n$ clusters, which leads to a 
monotonically increasing trend of $\rm{L_{tot}}$ and 
$ \rm{S_{tot}}$ in these clusters. The ${\langle\rm{L_M}\rangle}$ and 
${\langle\rm{L_{Pt}}\rangle}$ for Co$_{13}$Pt$_n$ clusters shows a 
trend similar to Fe$_{13}$Pt$_n$, i.e. with increasing number of 
Pt atoms, the ${\langle\rm{L_M}\rangle}$ (as defined in 
Table VI) on the Co$_{13}$ cluster 
decreases and the ${\langle\rm{L_{Pt}}\rangle}$ on the Pt atoms is enhanced.
The ${\langle\rm{S_M}\rangle}$ on Co$_{13}$ is merely constant, while it 
increases monotonically for Pt with more Pt atoms on the cluster surface.

\subsection{Ni$_{13}$Pt$_n$ clusters}
In this case, the $ \rm{L_{tot}}$ and $ \rm{S_{tot}}$ also show a regular 
increase with the number of Pt atoms. The ${\langle\rm{L_M}\rangle}$ 
and ${\langle\rm{L_{Pt}}\rangle}$ for this cluster do not much 
depend on the number of Pt atoms.
However, a decrease of ${\langle\rm{L_{Pt}}\rangle}$ and 
${\langle\rm{S_{Pt}}\rangle}$ from Ni$_{13}$Pt$_{5}$ to Ni$_{13}$Pt$_{20}$ 
cluster is also observed. This is probably caused by a 
structural instability occuring for this composition upon 
the relaxation. The geometry optimization of this cluster converges 
to a structure with different symmetry, with a ferromagnetic 
ordering, where the Ni atoms are placed closer to the
surface of the cluster as shown in Fig.\ 6. The reason for the 
segregation of Ni atoms towards the surface may be due to its lower 
surface energy compared to Pt~\cite{alonso-1927}. 
Another aspect related to this may be observed in 
the right panel of Figs. 8 and 9 showing the variation of onsite orbital 
and spin moments with respect to the distance. The large variations 
in orbital and spin moments just 
occur because of the structural distortion for this cluster composition. 
Comparing all three cases of capped clusters, we find that 
the presence of Pt atoms on M$_{13}$ affects the orientation 
of core orbital moments in such a way that they always prefer to be in  
non-collinear alignment for the M$_{13}$Pt$_n$ clusters, which is not 
the case in the uncapped M$_{13}$ clusters. On the other hand, directions 
of individual spin moments remain always collinear for the same clusters 
indicating that they are less affected by the Pt atoms. 

Regarding the MAE of capped M$_{13}$Pt$_n$ clusters, we observe that 
the symmetry rules the magnitude of the effect like for M$_{13}$ systems. 
For example, in Fe$_{13}$Pt$_4$ (see Fig. \ 10) with second-order anisotropy,
we find a high MAE value, even larger by a factor of two ($\sim$
7 meV/cluster) compared to JT-distorted Fe$_{13}$.

\begin{table}
\caption{\label{tab:table2} The orbital and spin moments for 
the binary Pt-M clusters in $\mu_B$/atom, where we have distinguished 
core and shell contributions. Cluster averaged values are defined for 
${\langle\rm{L_M}\rangle}$ = $\frac{1}{13}$$\sum_{i=1}^{13}{|\mathbf{L}_i|}$ and ${\langle\rm{S_M}\rangle}$ = $\frac{1}{13}$$\sum_{i=1}^{13}{|\mathbf{S}_i|}$ (M represents the 13-atom Fe, Co and Ni clusters). The averaged 
values for ${\langle\rm{L_{Pt}}\rangle}$ = $\frac{1}{n}$$\sum_{i=1}^{n}{|\mathbf{L}_i|}$ and ${\langle\rm{S_{Pt}}\rangle}$ = $\frac{1}{n}$$\sum_{i=1}^{n}{|\mathbf{S}_i|}$ ($n$ is the number of Pt atoms and here, $n$= 3, 5, 20).
$|\mathbf{L_{tot}|}$ and $|\mathbf{S_{tot}|}$ 
are the corresponding absolute values of the total orbital 
and spin moment for every composition of the binary clusters.\\ }
\begin{ruledtabular}
{\scriptsize
\begin{tabular}{lcccccr}
Cluster&${\langle\rm{L_M}\rangle}$&${\langle\rm{L_{Pt}}\rangle}$&${\langle\rm{S_M}\rangle}$&${\langle\rm{S_{Pt}}\rangle}$&$|\mathbf{L_{tot}|}$&$|\mathbf{S_{tot}|}$\\
\hline
Fe$_{13}$Pt$_3$& 0.09 & 0.08 & 2.86 & 0.36 & 1.43 & 38.3 \\
Fe$_{13}$Pt$_5$& 0.08 & 0.134 & 2.97 & 0.526 & 1.76 & 41.3 \\
Fe$_{13}$Pt$_{20}$& 0.05 & 0.17 & 2.18 & 0.401 & 2.74 & 36.4 \\  
\hline
Co$_{13}$Pt$_3$& 0.1 & 0.13 & 1.87 & 0.42 & 1.68 & 25.6 \\ 
Co$_{13}$Pt$_5$& 0.095 & 0.16 & 1.83 & 0.47 & 2.06 & 26.1 \\
Co$_{13}$Pt$_{20}$& 0.07 & 0.3 & 1.88 & 0.60 & 4.9 & 36.2 \\ 
\hline
Ni$_{13}$Pt$_3$& 0.06 & 0.2 & 0.81 & 0.402 & 1.61 & 11.7 \\
Ni$_{13}$Pt$_5$& 0.05 & 0.26 & 0.80 & 0.512 & 2.01 & 13.0 \\
Ni$_{13}$Pt$_{20}$& 0.08 & 0.2 & 0.81 & 0.36 & 3.55 & 17.6 \\
\end{tabular}
}
\end{ruledtabular}
\end{table}


\begin{figure*}
\includegraphics[scale=0.15]{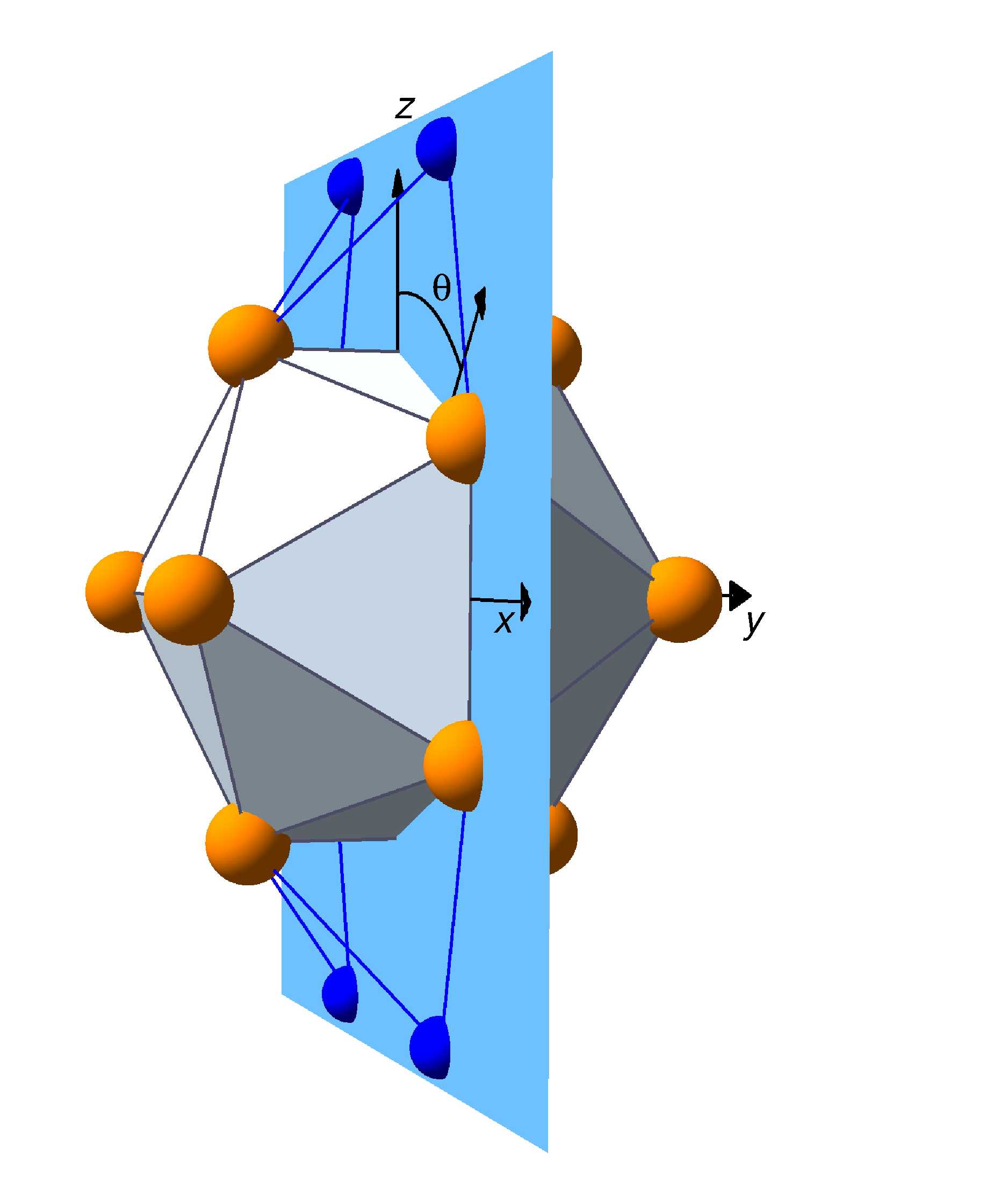}
\hspace{0.5cm} 
\includegraphics[scale=0.50]{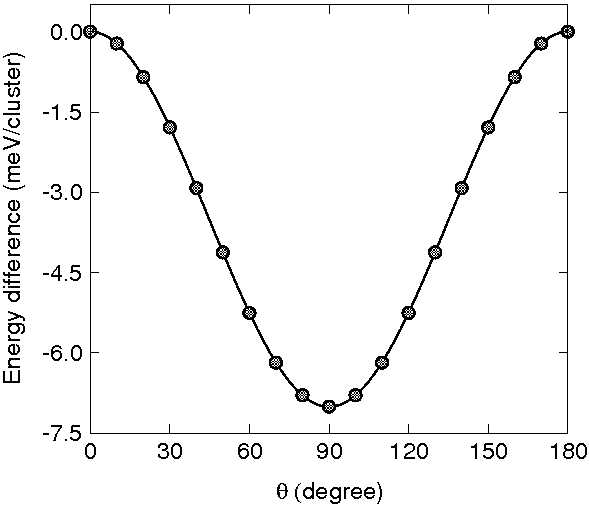} 
\caption{\label{fig:epsart}Left: The $x$-$z$ plane (where $\theta$ is 
varied) of the Fe$_{13}$Pt$_4$ cluster, showing the positions of 
the Pt-atoms relative to the ICO in the unrelaxed 
configuration. Light (orange) and dark (blue) spheres 
are the Fe and Pt atoms, respectively. Right: The $\theta$-dependent 
energy differences for the relaxed Fe$_{13}$Pt$_4$ cluster. The energy 
difference is in units of meV per cluster and is defined 
as $\Delta E= E-E_{\theta=0}$.}
\end{figure*}

\section{Conclusion}
The spin and orbital magnetic moments are calculated for 
M$_{13}$ and M$_{13}$Pt$_{n}$ clusters including spin-orbit interaction 
using DFT methods. The studies on MAE of M$_{13}$ clusters 
show that the calculated MAE values of relaxed M$_{13}$ clusters 
are considerably enhanced relative to both the ideal icosahedral 
clusters and to the 
corresponding bulk values in case of Fe$_{13}$ and Co$_{13}$. 
For Ni$_{13}$, the MAE is smaller than that of the bulk value of fcc Ni.
The MAEs for relaxed clusters are found to be affected by 
the degree of relaxation due to which 
Fe$_{13}$ clusters have a high anisotropy in comparison with 
Co$_{13}$ and Ni$_{13}$. Iron is a special case for which we 
obtain two local energy 
minima~\cite{rollmann-2007, rollmann-th-2007}, corresponding to 
the JT and MT 
cases. The JT-distorted Fe$_{13}$ cluster exhibits
an anisotropy approximately 5 times larger than the 
partially MT cluster due to the low symmetry 
in the former case. 
The present calculations of MAE agree well with the qualitative 
prediction of the Heisenberg model for the ${\theta}$-dependent 
energy differences. With respect to the spin and orbital 
moments, both the capped and the free clusters show an increased 
value of orbital and spin moments compared to the bulk. For the 
capped clusters, the spin moments on Pt atoms 
remain unaffected by the host atoms. 
Finally, we infer that deposited transition metal 
clusters, with very large effects of relaxations, may even exhibit 
larger MAE values. Self-assembly of such clusters, like in case 
of Fe-Pt~\cite{gruner-2008} and Co~\cite{xiao-2009}, may then approach 
the class of functional magnetic materials of use for magnetic storage devices.

\section*{Acknowledgments}
The authors would like to acknowledge Helmut Eschrig and Klaus 
Koepernik for helpful discussions. This work was 
supported by SFB 445 and SPP 1145.


\begin{thebibliography}{99}

\bibitem{polesya-2006}
S. Polesya, O. \v{S}ipr, S. Bornemann, J. Min\'{a}r, and H. Ebert, Europhys. Lett. {\bf 74}, 1074 (2006).

\bibitem{sipr-2007}
O. \v{S}ipr, S. Polesya, J. Min\'{a}r, and H. Ebert, J. Phys.: Condens. Matter. {\bf 19}, 446205 (2007).

\bibitem{xie-2002}
Y. Xie, and J. A. Blackman, Phys. Rev. B {\bf 66}, 085410 (2002).

\bibitem{dieguez-2001}
O. Di\`{e}guez, M. M. G. Alemay, C. Rey, P. Ordej\'{o}n, and L. J. Gallego, Phys. Rev. B {\bf 63}, 205407 (2001).

\bibitem{postnikov-2003}
A. V. Postnikov, P. Entel, and J. M. Soler, Eur. Phys. J. D {\bf 25}, 261 (2003).

\bibitem{duan-2001}
H. M. Duan and Q. Q. Zheng, Phys. Lett. A {\bf 280}, 333 (2001).

\bibitem{apsel-1996}
S. E. Apsel, J. W. Emmert, J. Deng, and L. A. Bloomfield, Phys. Rev. Lett. {\bf 76}, 1441 (1996).

\bibitem{winter-bovin-1991}
B. J. Winter, T. D. Klots, E. K. Parks, and S. J. Riley, Z. Phys. D {\bf 19}, 375 (1991);
J.-O. Bovin, and J.-O. Malm, Z. Phys. D {\bf 19}, 293 (1991). 

\bibitem{dunlap-1990}
B. I. Dunlap, Phys. Rev. A {\bf 41}, 5691 (1990).

\bibitem{tiago-2006}
M. L. Tiago, Y. Zhou, M. M. G. Alemany, Y. Saad, and J. R. Chelikowsky, Phys. Rev. Lett. {\bf 97}, 147201 (2006).

\bibitem{entel_pm-2008}
P. Entel, M. E. Gruner, G. Rollmann, A. Hucht, S. Sahoo, A. T. Zayak, H. C. Herper, A. Dannenberg,  Phil. Mag. {\bf 88}, 2725 (2008).

\bibitem{rollmann-2007}
G. Rollmann, M. E. Gruner, A. Hucht, R. Meyer, P. Entel, M. L. Tiago, and J. R. Chelikowsky, Phys. Rev. Lett. {\bf 99}, 083402 (2007).

\bibitem{gruner-2008}
M. E. Gruner, G. Rollmann, P. Entel, and M. Farle, Phys. Rev. Lett. {\bf 100}, 087203 (2008).

\bibitem{xiao-2009}
Xiao $\it{et al.}$, http://arxiv.org/abs/0906.4645.

\bibitem{futschek-2006}
T. Futschek, J. Hafner, and M. Marsman, J. Phys.: Condens. Matter, {\bf 18}, 9703 (2006).

\bibitem{sandra-1998-NC}
L. M. Sandratskii, Adv. Phys. {\bf 47}, 91 (1998).

\bibitem{sandra-1996-NC-SOI}
L. M. Sandratskii, and J. K\"{u}bler, Mod. Phys. Lett. B {\bf 10}, 189 (1996).

\bibitem{oda-1998}
T. Oda, A. Pasquarello, and R. Car, Phys. Rev. Lett. {\bf 80}, 3622 (1998).

\bibitem{hobbs-2000}
D. Hobbs, G. Kresse, and J. Hafner, Phys. Rev. B {\bf 62}, 11556 (2000).

\bibitem{lounis-2007}
S. Lounis, P. Mavropoulos, R. Zeller, P. Dederichs, and S. Bl\"{u}gel, Phys. Rev. B {\bf 75}, 174436 (2007).

\bibitem{strandberg-2007}
T. O. Strandberg, C. M. Canali, and A. H. MacDonald, Nature Mater. {\bf 6}, 648 (2007).

\bibitem{gambardella-2003}
P. Gambardella, S. Rusponi, M. Veronese, S. S. Dhesi, C. Grazioli, A. Dallmeyer, I. Cabria, R. Zeller, P. H. Dederichs, K. Kern, C. Carbone, and H. Brune, Science {\bf 300}, 1130 (2003).

\bibitem{luis-2006}
F. Luis, J. Bartolom\'e, M. J. Martinez, and L. M. Garcia, J. Appl. Phys. {\bf 99}, 08G705 (2006).

\bibitem{bansmann-2005}
J. Bansmann, S. H. Baker, C. Binns, J. A. Blackman, J.-P. Bucher, J. Dorantes-D\'{a}vila, V. Dupuis, L. Favre, D. Kechrakos, A. Kleibert, K.-H. Meiwes-Broer, G. M. Pastor, A. Perez, O. Toulemonde, K. N. Trohidou, J. Tuaillon, and Y. Xie, Surf. Sci. Rep. {\bf 56}, 189 (2005).

\bibitem{dennler-2004}
S. Dennler, J. Morillo, and G. M. Pastor, J. Phys.: Condens. Matter. {\bf 16}, S2263 (2004).

\bibitem{navia-2004}
M. Mu\~{n}oz-Navia, J. Dorantes-D\'{a}vila, and G. M. Pastor, J. Phys.: Condens. Matter. {\bf 16}, S2251 (2004).

\bibitem{masahito-2008}
M. Tsujikawa, A. Hosokawa, and T. Oda, Phys. Rev. B {\bf 77}, 054413 (2008).

\bibitem{medina-2003}
R. F\'elix-Medina, J. Dorantes-D\'avila, and G. M. Pastor, Phys. Rev. B {\bf 67}, 094430 (2003).
\bibitem{nonas-2001}
B. Nonas, I. Cabria, R. Zeller, and P. H. Dederichs, Phys. Rev. Lett. {\bf 86}, 2146 (2001).

\bibitem{andriotis-2004}
A. N. Andriotis, and M. Menon, Phys. Rev. Lett. {\bf 93}, 026402 (2004).

\bibitem{rohart-2006}
S. Rohart, C. Raufast, L. Favre, E. Bernstein, E. Bonet, and V. Dupuis, Phys. Rev. B {\bf 74}, 104408 (2006).

\bibitem{tsujikawa-2008}
M. Tsujikawa, A. Hosokawa, and T. Oda, Phys. Rev. B {\bf 77}, 054413 (2008).

\bibitem{pick-2004}
S. Pick, V. S. Stepanyuk, A. L. Klavsyuk, L. Niebergall, W. Hergert, J. Kirschner, and P. Bruno, Phys. Rev. B {\bf 70}, 224419 (2004).

\bibitem{pastor-1995}
G. M. Pastor, J. Dorantes-D\'avila, S. Pick, and H. Dreysse, Phys. Rev. Lett. {\bf 75}, 326 (1995).

\bibitem{ferrer-2007}
L. Fern\'andez-Seivane, and J. Ferrer, Phys. Rev. Lett. {\bf 99}, 183401 (2007).

\bibitem{hong-1991}
J. Hong, and R. Q. Wu, J. Appl. Phys. {\bf 69}, 6147 (1991).

\bibitem{fritsch-2008}
D. Fritsch, K. K\"{o}pernik, M. Richter, and H. Eschrig, J. Comp. Chem. {\bf 29}, 2210 (2008).

\bibitem{blonski-2009}
P. B\l o\'{n}ski and J. Hafner, Phys. Rev. B {\bf 79}, 224418 (2009).

\bibitem{guez-2001}
O. Di\'eguez, M. M. G. Alemany, C. Rey, P. Ordejon, and L. J. Gallego, Phys. Rev. B {\bf 63}, 205407 (2001).

\bibitem{kresse-cms}
G. Kresse, and J. Furthm\"{u}ller, Comput. Mater. Sci. {\bf 6}, 15 (1996), 

\bibitem{kresse-1996}
G. Kresse, and J. Furthm\"{u}ller, Phys. Rev. B {\bf 54}, 11169 (1996), 

\bibitem{kresse-1999}
G. Kresse, and D. Joubert, Phys. Rev. B {\bf 59}, 1758 (1999).

\bibitem{soler-2002}
J. M. Soler, E. Artacho, J. D. Gale, A. Gracia, J. Junquera, P. Ordejon, and D. Sanchez-Portal, J. Phys.: Condens. Matter {\bf 14}, 2745 (2002).

\bibitem{klaus-1999}
K. Koepernik and H. Eschrig, Phys. Rev. B {\bf 59}, 1743 (1999).

\bibitem{perdew-1991}
J. P. Perdew, in {\em Electronic Structure of Solids'91,} edited by 
P.~Ziesche and H.~Eschrig (Berlin, Akademie Verlag, 1991), pp. 11-20.

\bibitem{blochl1994}
P. E. Bl\"{o}chl, Phys. Rev. B {\bf  50}, 17953 (1994). 

\bibitem{perdew-1996}
J. P. Perdew, K. Burke, and M. Ernzerhof, Phys. Rev. Lett. {\bf 77}, 3865 (1996).

\bibitem{troullier-1991}
N. Troullier and J. L. Martins,  Phys. Rev. B {\bf 43}, 1993 (1991).

\bibitem{junquera-2001}
J. Junquera, O. Paz, D. Sanchez-Portal, and E. Artacho, Phys. Rev. B {\bf 64}, 235111 (2001).

\bibitem{fernandez-2006}
L Fernandez-Seivane, M. A. Oliveira, S. Sanvito and J. Ferrer, J. Phys.: Condens. Matter {\bf 18}, 7999 (2006).

\bibitem{FPLO}
http:/\!/www.fplo.de/

\bibitem{perdew-1992}
J. P. Perdew and Y. Wang, Phys. Rev. B {\bf 45}, 13244 (1992).

\bibitem{Eschrig:2004}
H.~Eschrig, M.~Richter, and I.~Opahle, in:
{\em Relativistic Solid State Calculations.}
{\em Relativistic Electronic Structure Theory -- Part II: Applications}, edited by P. Schwerdtfeger (Amsterdam: Elsevier, 2004), pp. 723-776.

\bibitem{rollmann-2004}
G. Rollmann, S. Sahoo, and P. Entel, Phys. Stat. Solidi (A) {\bf 201}, 3263 (2004).

\bibitem{rollmann-2006}
G. Rollmann, P. Entel, and S. Sahoo, Comput. Matter. Sci. {\bf 35}, 275 (2006).

\bibitem{rollmann-th-2007}
G. Rollmann, Ph.D. Thesis, University of Duisburg-Essen (2007).

\bibitem{mackay-1962}
A. L. Mackay, Acta Cryst. {\bf 15}, 916 (1962).

\bibitem{jahn-1937}
H. A. Jahn and E. Teller, Proc. R. Soc. London A {\bf 161}, 220 (1937).

\bibitem{entel-2008}
P. Entel, M. E. Gruner, A. Hucht, R. Meyer, G. Rollmann, S. Sahoo, and S. K. Nayak, in: {\em Mesoscopic, Nanoscopic and Macroscopic Materials}, edited by S. M. Bose, S. N. Behera, and B. K. Roul, AIP Conf. Proc. 1063, 3 (2008), pp. 3-17

\bibitem{guirado2003}
R. A. Guirado-L\'{o}pez, J. Dorantes-D\'{a}vila, and G. M. Pastor, Phys. Rev. Lett. {\bf 90}, 226402 (2003).

\bibitem{Gruner2008}
M. E. Gruner, P. Entel, I. Opahle, and M. Richter,
J. Mater. Sci. {\bf 43}, 3825 (2008).

\bibitem{yane-2009}
R Yanes, and O Chubykalo-Fesenko,
J. Phys. D: Appl. Phys. {\bf 42}, 055013 (2009).

\bibitem{halilov-1998}
S. V. Halilov, A. Ya. Perlov, P. M. Oppeneer, A. N. Yaresko, and V. N. Antonov, Phys. Rev. B {\bf 57}, 9557 (1998).


\bibitem{boyen-2004}
H. G. Boyen, K. Fauth, B. Stahl, P. Ziemann, G. K\"astle, F. Weigl, F. Banhart, M. Hessler, G. Sch\"utz, N. S. Gajbhiye, J. Ellrich, H. Hahn, M. B\"uttner, M. G. Garnier, and P. Oelhafen, in: Verhandl. DPG (VI), {\bf 39}, 1/ 253 (2004).

\bibitem{ebert-2006}
H. Ebert, S. Bornemann, J. Min\'{a}r, P. H. Dederichs, R. Zeller, 
and I. Cabria, Comput. Matter. Sci. {\bf 35}, 279 (2006).


\bibitem{alonso-1927}
J. A. Alonso and N. H. March, {\em Electrons in Metals and Alloys} 
(Academic, London, 1989), pp. 433-444.

\end{thebibliography}
\end{document}